\documentclass[%
reprint,
 amsmath,amssymb,
 aps,
]{revtex4-2}

\usepackage{graphicx}
\usepackage{dcolumn}
\usepackage{bm}
\usepackage{soul} 
\usepackage{xcolor} 
\sethlcolor{yellow} 

\begin{document}


\title{Photon-Number-Resolving Detector Based on a Cascade of Waveguide-Coupled Quantum Emitters}

\author{Abdolreza Pasharavesh}
 \email{Contact author: apashara@uwaterloo.ca}
 \author{Sai Sreesh Venuturumilli}%
\author{Michal Bajcsy}%
\affiliation{%
 Institute for Quantum Computing, University of Waterloo, Waterloo, ON, Canada\\ and Department of Electrical and Computer Engineering, University of Waterloo, Waterloo, ON, Canada
}%

\begin{abstract}
We investigate the operation of  a photon-number-resolving (PNR) detector consisting of a cascade of waveguide-coupled $\Lambda$-type emitters, where each waveguide-coupled emitter extracts a single photon from the input light and sends it to a single-photon detector. Using Green’s function and input-output formalisms, we derive the scattering matrices and photon-photon correlators for individual scatterers. By cascading these results, we obtain a closed-form expression for the detector's precision in the linear regime and predict how correlations generated by nonlinear photon-photon interactions influence this precision. To evaluate the performance of this  PNR detector in the nonlinear regime, we apply the quantum trajectory method to the cascaded setup, calculating the achievable precision and analyzing its dependence on key system parameters, such as the number of emitters and their coupling strength to the waveguide. We compare the performance of the proposed PNR detector with that of a conventional PNR scheme based on spatial demultiplexing via beamsplitters. Our results indicate that the proposed scheme can outperform conventional detectors under realistic conditions, making it a promising candidate for next-generation PNR detection.
\end{abstract}

\maketitle

\onecolumngrid

\section{\label{sec:level1}Introduction}

The development of single-photon sources and single-photon detectors enabled experimental studies of quantum light sources at the single-photon level, and such sources have proved a critical tool in a variety of applications, such as quantum communication \cite{RN180}, information \cite{RN181, RN184}, sensing \cite{RN181} and imaging \cite{RN182}.  At the same time, quantum states of light with multiple photons exist in a higher-dimensional space and allow for even more applications in these areas \cite{RN188, RN187, RN194}. Thus, generation and robust detection of multi-photon states open experimental access to study the rich space of non-classical states of light by granting new ways to study fundamental aspects of quantum optics. PNR detection is required to characterize and diagnose many-photon states of light by measuring their photon-number distributions. Furthermore, generalized detection schemes composed of such detectors can lend access to novel forms of quantum tomography.  In particular, characterizing complex, correlated many-photon states \cite{RN202, RN173, RN203, RN114, RN204} can help to reveal dynamics of many-body quantum systems. 

Deterministic photon subtraction, where exactly one photon is removed from the input field has been experimentally implemented using the single-photon Raman interaction (SPRINT) \cite{RN114, RN171, RN36} in a $\Lambda$-type emitter directionally coupled to a waveguide \cite{RN36}, as well as via Rydberg blockade in atomic ensembles \cite{RN190}. In the former method, a single photon from the incoming field is re-routed by changing the state of a chirally coupled $\Lambda$-type emitter, after which the emitter no longer interacts with the remaining incoming light. Therefore, a single-photon worth of energy is removed from the input light field. When the re-routed single-photon is detected, a given single-photon subtracted state of light is heralded in the outgoing field. In the strong coupling regime, we can deterministically remove the first arriving photon from an arbitrary input field, resulting in a temporally entangled state between the subtracted photon and the remaining light pulse. A particular choice of measurement was shown to reveal single-moded non-Gaussian light \cite{RN171}. Also, the entangled temporal mode structure of this state can hold utility in areas of quantum information utilizing resources of correlated multi-partite states \cite{RN186, RN185}. Furthermore, cascading SPRINT systems \cite{RN114} can extend the class of multi-photon states that can be generated \cite{RN229} and has in the past been suggested for  photon-number-resolved (PNR) detection \cite{RN98}.

Multiple-photon propagation in waveguides through quantum emitters have been experimentally \cite{RN202, RN207, RN208, RN210, RN204} and theoretically \cite{RN219, RN221, RN220, RN213, RN163} studied in a variety of settings. Already, the relatively simple case of two photons propagating through a single two-level emitter provides novel output light with features of squeezing, entanglement, and tuneable bunching \cite{RN209, RN211, RN208, RN210, RN212}. More generally, many-photon transport through many two-level and multi-level quantum emitters in a waveguide has shown to output highly non-classical, strongly correlated states and can showcase novel many-body phases of light \cite{RN218, RN202, RN173, RN217, RN216}. A suite of theoretical techniques has been developed to study quantum dynamics of multi-photon quantum light transport through quantum emitters. These include Master equation-based input-output theory \cite{RN162}, scattering matrix formalisms \cite{RN215, RN213, RN214}, Green’s function \cite{RN218} and matrix product state-based approaches \cite{RN218, RN163}, among others \cite{RN217, RN216}.

In this paper, we develop a theoretical formalism to study multi-photon transport through a cascade of SPRINT-based subtractors – with a focus on studying photon-number-resolving capability of the system. This includes analyzing the single- and multi-photon scattering characteristics of individual emitter-based subtractors \cite{RN173} in the time domain and combining them into emitter cascades to evaluate the precision of the proposed detectors. We examine photon propagation in two regimes: linear and nonlinear. In the linear regime, multi-photon propagation can be treated as a series of independent single-photon events, allowing single-photon transport knowledge to infer multi-photon behavior. In the nonlinear regime, photons can arrive within the emitter’s lifetime, leading to correlations between input photons due to emitter-mediated photon-photon interactions. The results are analyzed and compared in both regimes and are verified through numerical simulations using the SLH framework of input-output theory combined with the quantum trajectory method.

\section{Modeling and formulation}
\subsection{Cascaded photon subtraction scheme}
Fig. 1.a shows the schematic of the waveguide-QED system used as deterministic single-photon subtractor. The setup includes a single three-level $\Lambda$-type quantum emitter (atom) coupled to two waveguides. The waveguides are chiral \cite{RN110}, allowing light to propagate in one direction only, as indicated by the arrows. Waveguide 1, depicted in red, is coupled to the $\left| 1 \right\rangle  - \left| 3 \right\rangle $ transition, while waveguide 2, shown in blue, is coupled to the $\left| 2 \right\rangle  - \left| 3 \right\rangle $ transition. These waveguides can be either physically separate or distinct modes of a single waveguide such as counter propagating modes of a bidirectional chiral waveguide \cite{RN137}. We use ${a_{i,\omega }}$ to denote the annihilation operator for an excitation of guided mode $i$ at frequency $\omega $. Additionally, ${\gamma _g}$ characterizes the coupling strength and defines the rate at which the emitter emits and absorbs photons to and from the guided modes. Photon subtraction occurs via a single-photon Raman interaction, a coherent process during which the $\Lambda$-type atom transitions from one ground state to another by absorbing a photon from one light mode and emitting another photon into a different mode \cite{RN44}. In the configuration shown, the atom is initially in state $\left| 1 \right\rangle $, and the input pulse is introduced into the input mode of waveguide 1. Consequently, the subtracted photon exits the subtractor through the output mode of waveguide 2, while the remaining photons exit via the output of waveguide 1. If the input photons are separated by a duration longer than the atom's absorption/emission time, only a single photon will interact with the atom at any given time, resulting in no atom-mediated photon-photon interaction. In this case, the probability of each photon being subtracted can be calculated independently based on its frequency content and the system's single-photon transfer function \cite{RN170}. After one photon is subtracted, the atom in state $\left| 2 \right\rangle $ and no longer interacts with photons in waveguide 1, so any subsequent photons are directly transferred to the output mode of waveguide 1. When two photons arrive at the atom within a time interval shorter than the atomic response time, they can activate the system’s nonlinearity by saturating the atom.

\begin{figure}
\includegraphics{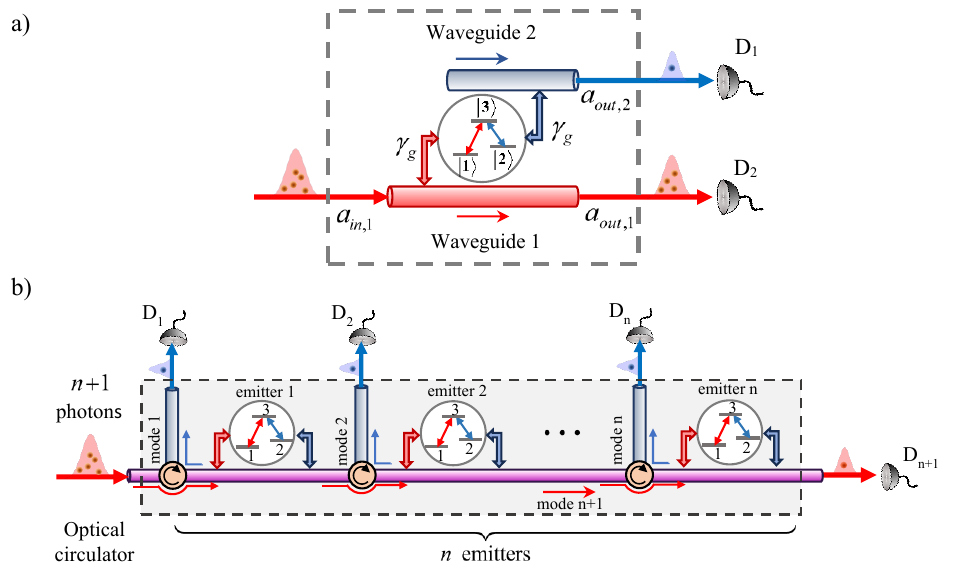}
\caption{\label{fig:epsart} a) Schematic of the photon subtractor as a three-port quantum optics device consisting of a $\Lambda$-type emitter coupled to two chiral (one-way) waveguide modes, and b) A cascade of $n$ photon subtractors forming a PNR detector capable of resolving up to $n + 1$ photons.
}
\end{figure}

If two single-photon detectors are placed at the outputs of the waveguide modes, as shown in the figure, the single-photon subtractor can function as a PNR detector, capable of counting photons in pulses with up to two photons. In this configuration, the number of detectors registering a click corresponds to the number of photons contained in the input pulse. Following this idea further, PNR detectors capable of resolving higher photon counts can be constructed by cascading single photon subtractors over a shared waveguide, as illustrated in Fig. 1.b. In this setup, the atoms each coherently absorb one photon from the right-propagating red mode and divert those photons to the left-propagating blue mode. Optical circulators \cite{RN161} placed between the emitters extract the single photons subtracted by each emitter and send them to separate detectors via the vertical blue waveguides. In this way, a cascade of $n$ emitters can resolve light pulses containing up to $n + 1$ photons. Since there is no external photon input to the waveguide modes $1$ to $n$, they can only be in either a vacuum state or contain a single photon routed by their corresponding circulator from the left propagating field of the chiral waveguide. In contrast, the chiral waveguide mode can contain any number of photons, depending on the photon count in the input pulse. Therefore, the waveguides Hilbert spaces will be ${\cal H}_{{\rm{wg}}}^{i = 1, \ldots ,n} = {\rm{span}}\left\{ {\left| {{\rm{vac}}} \right\rangle ,{{\left| t \right\rangle }_i}} \right\}$ and  ${\cal H}_{{\rm{wg}}}^{i = n + 1} = {\rm{span}}\left\{ {\left| {{\rm{vac}}} \right\rangle ,{{\left| {{t_1},{t_2}, \cdots ,{t_m}} \right\rangle }_i}} \right\}$, where:
\begin{eqnarray}
  \begin{array}{l}
  {{\left| {{t}_{1}},{{t}_{2}},\cdots ,{{t}_{m}} \right\rangle }_{i}}=\prod\limits_{n=1}^{m}{a_{i,{{t}_{n}}}^{\dagger }\left| \text{vac} \right\rangle }\,\,\,\,\,\,\,\,\forall m\in \mathbb{N} \\ 
 {{\left| t \right\rangle }_{i}}=a_{i,t}^{\dagger }\left| \text{vac} \right\rangle
 \end{array}
\end{eqnarray}

In the above expression, $a_{i,t}^\dag  = \int {a_{i,\omega }^\dag \exp \left( {i\omega t} \right){{d\omega } \mathord{\left/
 {\vphantom {{d\omega } {\sqrt {2\pi } }}} \right.
 \kern-\nulldelimiterspace} {\sqrt {2\pi } }}}$ is the temporal creation operator creating a photon in the waveguide mode $i$ at time $t$ and in the atom’s position. These photons propagate freely in the respective waveguides before and after interacting with the atom. The total Hilbert space of the composite system can be written as the tensor product of these Hilbert spaces with the atomic Hilbert spaces ${\cal H}_{{\rm{atom}}}^{i = 1, \ldots ,n} = {\rm{span}}\left\{ {{{\left| 1 \right\rangle }_i},{{\left| 2 \right\rangle }_i},{{\left| 3 \right\rangle }_i}} \right\}$. Therefore, a basis for the Hilbert space of the entire system can be represented as:
\begin{eqnarray}
    \left[ {\mathop  \otimes \limits_{i = 1}^n {{\left| \sigma  \right\rangle }_i}} \right] \otimes \left[ {\mathop  \otimes \limits_{i = 1}^n {{\left| t \right\rangle }_i}} \right] \otimes {\left| {{t_1},{t_2}, \cdots ,{t_m}} \right\rangle _{n + 1}}
\end{eqnarray}
with $\sigma  = 1,2,3$ denoting the corresponding atomic level of the emitters. If either of the waveguide modes is in the vacuum state, the corresponding element is replaced by "vac".

\subsection{Scattering matrix calculation}

In quantum field theory, the effect of a local quantum system on propagating quantum fields coupled to it can be studied using the scattering matrix ${\bf{\Sigma }}$. It is essentially the interaction picture propagation operator of the system between two specific time points: long before the interaction starts ($t = {\tau ^ - } \to  - \infty $) and long after it ends ($t = {\tau ^ + } \to  + \infty $). According to this definition, the scattering matrix is given by ${\bf{\Sigma }} = {\cal T}\exp \left( { - i{H_I}\left( {{\tau ^ + } - {\tau ^ - }} \right)} \right)$, where  is the interaction picture Hamiltonian, and  represents the time-ordering operator. Different elements of the scattering matrix represent the transition amplitude from various input states of the system onto its distinct output states. For instance, consider a PNR detector consisting of $n$ emitters. It can successfully count the number of photons in an $n + 1$-photon pulse if one photon is detected at each output port and all emitters end up in state $\left| 2 \right\rangle $. This outcome corresponds to the following element of the scattering matrix:
\begin{eqnarray}
	\left[ {\mathop  \otimes \limits_{i = 1}^n {{\left\langle 2 \right|}_i}} \right] \otimes \left[ {\mathop  \otimes \limits_{i = 1}^{n + 1} {{\left\langle {{{t'}_i}} \right|}_i}} \right]{\bf{\Sigma }}\left[ {\mathop  \otimes \limits_{i = 1}^n {{\left| {{\rm{vac}}} \right\rangle }_i}} \right] \otimes {\left| {{t_1},{t_2}, \cdots ,{t_{n + 1}}} \right\rangle _{n + 1}} \otimes \left[ {\mathop  \otimes \limits_{i = 1}^n {{\left| 1 \right\rangle }_i}} \right]
\end{eqnarray}

In the above notation, ${t_1}$ to ${t_{n + 1}}$ are the time points at which the input photons are introduced to the detector's input port, while ${t'_1}$ to  represent the times at which the output photons are detected at their respective output ports. This means that $n$ photons have interacted with the emitters and been diverted to a different waveguide mode, while one photon has either not interacted with any of the emitters or has been absorbed and then re-emitted into the same waveguide mode. 

In a cascaded quantum system, the scattering matrix for the entire system can be expressed as a product of the scattering matrices of each individual cascaded element, integrated over all possible intermediate states. Furthermore, when all cascaded elements are of the same type, the scattering behavior of a single element can provide insight into how optical properties are altered as they propagate through the cascade. Therefore, we begin our analysis by focusing on the scattering behavior of a single photon subtractor, as illustrated in Fig. 1(a). When $N$ photons are introduced to the single photon subtractor, two distinct outcomes are possible: either one photon (e.g., $j$-th photon) is subtracted and diverted to the second waveguide mode, or no photon is subtracted. In the former case, the atom's state changes to $\left| 2 \right\rangle $, while in the latter case, the atom's state remains unchanged. These different outcomes correspond to distinct elements of the $N$-photon scattering matrix, denoted by $\Sigma _j^N$ and $\Sigma _0^N$, respectively:
\begin{eqnarray}
	\begin{array}{l}
\Sigma _j^N = \left\langle 2 \right| \otimes \left\langle {{{t'}_{j'}}} \right| \otimes \left\langle {{{t'}_1}, \ldots ,{{t'}_{j' - 1}},{{t'}_{j' + 1}}, \ldots ,{{t'}_N}} \right|{\bf{\Sigma }}\left| {{t_1}, \ldots ,{t_N}} \right\rangle  \otimes \left| {{\rm{vac}}} \right\rangle  \otimes \left| 1 \right\rangle \\
\Sigma _0^N = \left\langle 1 \right| \otimes \left\langle {{\rm{vac}}} \right| \otimes \left\langle {{{t'}_1}, \ldots ,{{t'}_N}} \right|{\bf{\Sigma }}\left| {{t_1}, \ldots ,{t_N}} \right\rangle  \otimes \left| {{\rm{vac}}} \right\rangle  \otimes \left| 1 \right\rangle 
\end{array}
\end{eqnarray}

We assume that photon arrival times at both the input and output are ordered such that ${t_1} \le {t_2} \le  \ldots  \le {t_N}$ and ${t'_1} \le {t'_2} \le  \ldots  \le {t'_N}$. Since the atom can absorb no more than one photon at a time, it is impossible to detect two consecutive photons at the outputs without sending any photons to the input, meaning that ${t_1} \le {t'_1} \le {t_2} \le {t'_2} \le  \ldots  \le {t_N} \le {t'_N}$. Additionally, since after the first photon is subtracted the atom no longer interacts with photons in the red waveguide, we have ${t'_i} = {t_i}$ for all photons sent and received after the photon subtraction occurs.

To calculate the matrix elements given in equation (4), we follow the approach proposed by Xu and Fan in Ref. \cite{RN162}. That work investigated multi-photon transport through a waveguide-coupled low-dimensional quantum system, establishing a connection between the general $N$-photon transport scattering matrix and the Green's functions of the local quantum system. The key significance of this work was that the Green's function could, in turn, be computed using an effective Hamiltonian that depends solely on the internal degrees of freedom of the local system. Later, Xu and Fan’s method was extended to time-dependent quantum systems coupled to multiple waveguides in Ref. \cite{RN163}. There, Trivedi et al. express the system’s propagator elements as functions of the Green's function  ${\cal G}_{{\tau ^ + },{\tau ^ - }}^{\sigma ',\sigma }$, which correlates the times of absorption and re-emission of photons to the external optical modes. In this notation, $\sigma $ and $\sigma '$ are states of the local quantum system before and after the interaction, respectively. For the case of $k$ photons interacting with the emitter, we have the following Green’s functions:
\begin{eqnarray}
\begin{array}{l}
{\cal G}_{{\tau ^ + },{\tau ^ - }}^{2,1}\left( {{t_1}, \ldots ,{t_k};{{t'}_1}, \ldots ,{{t'}_k}} \right) = \left( {\theta \left( {{{t'}_1} - {t_1}} \right)\prod\limits_{i = 2}^k {\theta \left( {{{t'}_i} - {t_i}} \right)\theta \left( {{t_i} - {{t'}_{i - 1}}} \right)} } \right)\gamma _g^k\left\langle 2 \right|{{\tilde \sigma }_{23}}\left( {{{t'}_k}} \right){{\tilde \sigma }_{31}}\left( {{t_k}} \right){\cal T}\left[ {\prod\limits_{i = 1}^{k - 1} {{{\tilde \sigma }_{13}}\left( {{{t'}_i}} \right){{\tilde \sigma }_{31}}\left( {{t_i}} \right)} } \right]\left| 1 \right\rangle \\
{\cal G}_{{\tau ^ + },{\tau ^ - }}^{1,1}\left( {{t_1}, \ldots ,{t_k};{{t'}_1}, \ldots ,{{t'}_k}} \right) = \left( {\theta \left( {{{t'}_1} - {t_1}} \right)\prod\limits_{i = 2}^k {\theta \left( {{{t'}_i} - {t_i}} \right)\theta \left( {{t_i} - {{t'}_{i - 1}}} \right)} } \right)\gamma _g^k\left\langle 1 \right|{\cal T}\left[ {\prod\limits_{i = 1}^k {{{\tilde \sigma }_{13}}\left( {{{t'}_i}} \right){{\tilde \sigma }_{31}}\left( {{t_i}} \right)} } \right]\left| 1 \right\rangle 
\end{array}
\end{eqnarray}
where $\theta \left( t \right)$ is the Heaviside step function. In the above equation, ${\cal G}_{{\tau ^ + },{\tau ^ - }}^{2,1}$ represents the case where the $k$-th arriving photon is successfully subtracted, while ${\cal G}_{{\tau ^ + },{\tau ^ - }}^{1,1}$ corresponds to the case where no photon has been subtracted and all photons exit through the red waveguide output. It should be noted that, since the emitter no longer interacts with input photons after subtracting a photon, the subtracted photon, if any, must be the last photon to interact with the emitter (i.e., the $k$-th photon in the case of $k$ interacting photons). Additionally, ${\tilde \sigma _{ij}}\left( t \right)$ are atomic transition operators that evolve in time under an effective Hamiltonian ${H_{{\rm{eff}}}}$ according to ${\dot \tilde \sigma _{ij}} = i[{H_{{\rm{eff}}}},{\tilde \sigma _{ij}}]$, starting from their Schrodinger picture counterparts ${\sigma _{ij}} = \left| i \right\rangle \left\langle j \right|$. The effective Hamiltonian is ${H_{{\rm{eff}}}} = {H_{{\rm{atom}}}} - \frac{i}{2}\sum\limits_{\mu  = 1}^2 {L_\mu ^\dag {L_\mu }}$ where ${H_{{\rm{atom}}}} = {\omega _{13}}{\sigma _{33}} + \left( {{\omega _{13}} - {\omega _{23}}} \right){\sigma _{22}}$, with ${\omega _{12}}$ and ${\omega _{23}}$ representing the atomic transition frequencies from ground states $\left| 1 \right\rangle $ and $\left| 2 \right\rangle $ to the excited state $\left| 3 \right\rangle $, respectively. Additionally, ${L_1} = \sqrt {{\gamma _g}} {\sigma _{13}}$ and ${L_2} = \sqrt {{\gamma _g}} {\sigma _{23}},$ are the coupling operators that couple the atom to the waveguide modes. Therefore, for ${\tilde \sigma _{13}}$ and ${\tilde \sigma _{23}}$ we have:
\begin{eqnarray}
	{\tilde \sigma _{13}}\left( t \right) = \exp \left( { - \left( {i{\omega _{13}} + {\gamma _g}} \right)t} \right){\sigma _{13}}\,\,\,\,\,\,\,\,\,\,\,\,{\tilde \sigma _{23}}\left( t \right) = \exp \left( { - \left( {i{\omega _{23}} + {\gamma _g}} \right)t} \right){\sigma _{23}}
    \end{eqnarray}
    
Substituting these operators into equation (5), gives the following Green’s functions:
\begin{eqnarray}
\begin{array}{l}
{\cal G}_{{\tau ^ + },{\tau ^ - }}^{2,1}\left( {{t_1}, \ldots ,{t_k};{{t'}_1}, \ldots ,{{t'}_k}} \right) = \left( {\theta \left( {{{t'}_1} - {t_1}} \right)\prod\limits_{i = 2}^k {\theta \left( {{{t'}_i} - {t_i}} \right)\theta \left( {{t_i} - {{t'}_{i - 1}}} \right)} } \right)\gamma _g^k\exp \left( {i{\omega _{12}}{{t'}_k}} \right)\exp \left( {\left( {i{\omega _{13}} + {\gamma _g}} \right)\sum\limits_{i = 1}^k {\left( {{t_i} - {{t'}_i}} \right)} } \right)\\
{\cal G}_{{\tau ^ + },{\tau ^ - }}^{1,1}\left( {{t_1}, \ldots ,{t_k};{{t'}_1}, \ldots ,{{t'}_k}} \right) = \left( {\theta \left( {{{t'}_1} - {t_1}} \right)\prod\limits_{i = 2}^k {\theta \left( {{{t'}_i} - {t_i}} \right)\theta \left( {{t_i} - {{t'}_{i - 1}}} \right)} } \right)\gamma _g^k\exp \left( {\left( {i{\omega _{13}} + {\gamma _g}} \right)\sum\limits_{i = 1}^k {\left( {{t_i} - {{t'}_i}} \right)} } \right)
\end{array}
\end{eqnarray}

Each input photon to the photon subtractor can have one of three outcomes: it may be absorbed by the emitter and diverted to the second waveguide mode, absorbed and re-emitted into the same mode, or pass through the atom without interacting with it. Based on this, and utilizing the Green's functions mentioned earlier, we can express the scattering matrix elements introduced in equation (4). For the case of one-photon transport, the matrix elements, after transforming to a rotating frame at the atomic transition frequencies, are given by:
\begin{eqnarray}
\begin{array}{l}
\Sigma _1^1\left( {{{t'}_1} - {t_1}} \right) =  - {\gamma _g}\exp \left( {{\gamma _g}\left( {{t_1} - {{t'}_1}} \right)} \right)\\
\Sigma _0^1\left( {{{t'}_1} - {t_1}} \right) = \delta \left( {{t_1} - {{t'}_1}} \right) - {\gamma _g}\exp \left( {{\gamma _g}\left( {{t_1} - {{t'}_1}} \right)} \right)
\end{array}
\end{eqnarray}

Additionally, for $N$-photon transport, the scattering matrix elements can be expressed as a product of the single-photon elements given above, with an additional multi-photon kernel (MPK) that accounts for photon reordering due to saturation effects:
\begin{eqnarray}
\Sigma _j^N\left( {{t_1}, \ldots ,{t_N};{{t'}_1}, \ldots ,{{t'}_N}} \right) = \left( {\left[ {\prod\limits_{i = 1}^{j - 1} {\Sigma _0^1\left( {{{t'}_i} - {t_i}} \right)} } \right]\Sigma _1^1\left( {{{t'}_j} - {t_j}} \right) + {\rm{MPK}}} \right)\left[ {\prod\limits_{i = j + 1}^N {\delta \left( {{{t'}_i} - {t_i}} \right)} } \right]
\end{eqnarray}
where:
\begin{eqnarray}
{\rm{MPK}} = \sum\limits_{{A_j}} {\left[ \begin{array}{l}
\left[ {\prod\limits_{\begin{array}{*{20}{c}}
{i = 1}\\
{{A_j}\left( i \right) = 1}
\end{array}}^{\left| {{A_j}} \right|} {\Sigma _{{\delta _{ij}}}^1\left( {{{t'}_i} - {t_i}} \right)} } \right] \times \\
\left[ {\prod\limits_{\begin{array}{*{20}{c}}
{i = 1}\\
{{A_j}\left( i \right) > 1}
\end{array}}^{\left| {{A_j}} \right|} {\left[ { - {\gamma _g}\left( {\prod\limits_{k = 1}^{{A_j}\left( i \right) - 1} {\delta \left( {{t_{B\left( {i - 1} \right) + k + 1}} - {{t'}_{B\left( {i - 1} \right) + k}}} \right)} } \right)\exp \left( {{\gamma _g}\left( {{t_{B\left( {i - 1} \right) + 1}} - {{t'}_{B\left( {i - 1} \right) + {A_j}\left( i \right)}}} \right)} \right)} \right]} } \right]
\end{array} \right]}\nonumber
\end{eqnarray}
\begin{eqnarray}
\end{eqnarray}

In the above expression, ${\delta _{ij}}$ is the Kronecker delta function,  is an ordered set of fewer than $j$ natural numbers whose sum is $j$, $\left| {{A_j}} \right|$ denotes its cardinal, and $B$ is a set whose elements are the partial sums of the first $i$ elements of ${A_j}$, i.e., $B\left( i \right)=\sum\nolimits_{k=1}^{i}{{{A}_{j}}\left( k \right)}$. For the case where no photon is subtracted, photon-photon interactions involving up to $N$ photons are possible. Therefore, in the above equation, $j$ should be replaced by $N$, and ${\delta _{ij}}$ should be set to zero. It is worth noting again that the expressions given in equations (9) and (10) are valid only when the input and output photon arrival times are sorted, and there is exactly one output photon between each pair of input photons (i.e., ${t_1} \le {t'_1} \le {t_2} \le {t'_2} \le  \ldots  \le {t_N} \le {t'_N}$), otherwise, the transition amplitude is zero.

\subsection{Equivalent input-output network model}

The proposed PNR detector in Fig. 1.b functions as a quantum input-output network (QION) with one input port and $n + 1$ output ports. By neglecting the propagation time of light between the emitters, an equivalent Markovian model for the network can be derived, comprising a higher-order local quantum system coupled to multiple field modes. This is achieved by eliminating the interconnections within the network and calculating an equivalent Hamiltonian along with the coupling operators responsible for the whole network's interaction with the external field modes. The SLH formalism, a framework based on quantum input-output theory, enables this process in a systematic way \cite{RN62, RN61}. In the SLH method, each node of a quantum network is modeled using a triple of operators $G = \left( {{\bf{S}},{\bf{L}},H} \right)$. In this notation, ${\bf{S}}$ is the scattering matrix, which describes how input fields are directly redistributed to output fields without being absorbed or re-emitted by the quantum system. Additionally, ${\bf{L}}$ is a column vector of coupling operators, representing the interaction between the internal degrees of freedom of the quantum system and the external quantum fields. Lastly, $H$ is the Hamiltonian of the quantum system, which governs its internal dynamics. It is important to note that the scattering matrix ${\bf{S}}$ differs from the one designated by ${\bf{\Sigma }}$ in the previous section, as the former accounts solely for the direct rerouting of fields without any interaction with the quantum system. When two individual quantum components are connected to form a quantum network, it becomes possible to replace both nodes with a single node characterized by an equivalent SLH triple. The equivalent operators can be determined using the composition rules outlined in the SLH formalism, which are derived from the governing equations of the composite system in the Heisenberg picture \cite{RN60}. For instance, suppose that two individual components or sub-networks, represented by ${G_1} = \left( {{{\bf{S}}_1},{{\bf{L}}_1},{H_1}} \right)$ and ${G_2} = \left( {{{\bf{S}}_2},{{\bf{L}}_2},{H_2}} \right)$, are cascaded with the output field of ${G_1}$ routed to the input of ${G_2}$. According to the cascade rule in the SLH framework, the SLH representation of the resulting composite node, referred to as the series product of the two nodes \cite{RN61}, will be ${G_T} = {G_1} \triangleright {G_2} = \left( {{{\bf{S}}_T},{{\bf{L}}_T},{{\bf{H}}_T}} \right)$ where ${{\bf{S}}_T} = {{\bf{S}}_2}{{\bf{S}}_1}$, ${{\bf{L}}_T} = {{\bf{L}}_2} + {{\bf{S}}_2}{{\bf{L}}_1}$, and ${H_T} = {H_1} + {H_2} - i{{\left( {{\bf{L}}_2^\dag {{\bf{S}}_2}{{\bf{L}}_1} - {\bf{L}}_1^\dag {\bf{S}}_2^\dag {{\bf{L}}_2}} \right)} \mathord{\left/
 {\vphantom {{\left( {{\bf{L}}_2^\dag {{\bf{S}}_2}{{\bf{L}}_1} - {\bf{L}}_1^\dag {\bf{S}}_2^\dag {{\bf{L}}_2}} \right)} 2}} \right.
 \kern-\nulldelimiterspace} 2}$ .
In the scheme shown in Fig. 1.b, the circulators directly transmit the right-propagating mode and route the single photons subtracted by the emitters (located to their right) to the vertical waveguide mode. Therefore, when deriving our input-output model, we can eliminate the circulators and assume that each emitter is directly coupled to the vertical waveguide to its left. By doing so, the network consists of $n$ emitters cascaded along a single guided mode (the right-propagating mode). For the $i$th emitter we have:
\begin{eqnarray}
{G_i} = \left( {{\bf{I}},{{\left[ {{L_1}, \ldots ,{L_{n + 1}}} \right]}^T},{H^{\left( i \right)}} = {\omega _{13}}\sigma _[45]^{\left( i \right)} + \left( {{\omega _{13}} - {\omega _{23}}} \right)\sigma _{22}^{\left( i \right)}\,} \right)\,\,\,\,\,\,\,{\rm{where:}}\,\,\,{L_j} = \sqrt {{\gamma _g}} \left( {\sigma _{23}^{\left( i \right)}{\delta _{ij}} + \sigma _{13}^{\left( i \right)}{\delta _{\left( {n + 1} \right)j}}} \right)
\end{eqnarray}

In the above expression, $\sigma _{kl}^{\left( i \right)} = \left| k \right\rangle \left\langle l \right|$ is the transition operator of the $i$-th emitter. Applying the cascade rule described above to the SLH form given in equation (11), the total SLH representation for the PNR detector is found as follows:
\begin{eqnarray}
{G_T} = {G_1} \triangleright  \ldots  \triangleright {G_n} = \left( {{\bf{I}},\sqrt {{\gamma _g}} {{\left[ {\sigma _{23}^{\left( 1 \right)}, \ldots ,\sigma _{23}^{\left( n \right)},\sum\limits_{i = 1}^n {\sigma _{13}^{\left( i \right)}} } \right]}^T},\left[ {\sum\limits_{i = 1}^n {{H^{\left( i \right)}}} } \right] + \frac{{{\gamma _g}}}{{2i}}\left[ {\sum\limits_{j = 2}^n {\left( {\sigma _{31}^{\left( j \right)}\sum\limits_{i = 1}^{j - 1} {\sigma _{13}^{\left( i \right)}} } \right)}  - {\rm{H}}{\rm{.c}}.} \right]\,} \right)
\end{eqnarray}
where ${\rm{H}}{\rm{.c}}.$ stands for Hermitian conjugate terms. To investigate the response of the atomic cascade to propagating light pulses, a modulating filter that generates the intended nonclassical field must also be cascaded to its input \cite{RN164}. With vacuum as input, the modulator should be capable of generating the desired nonclassical field. One possible realization of such a modulator is a cavity mode, where the coupling to the photon subtractor is defined according to the intended pulse profile. The statistics of the modulator's output is determined by the initial state of the cavity. If the input light to the photon subtractor has a multi-temporal mode structure, a separate cavity mode is used for each temporal mode of the input \cite{RN63, RN171}.

\section{Results and discussion}

In this section, we apply the mathematical models developed in previous section to examine the behavior of single- and multi-photon subtractors when subjected to input fields containing multiple photons. The quantum state of the $N$-photon input light is expressed in the following general form, where the photons are labeled based on their arrival times:
\begin{eqnarray}
	\left| {{\psi _{in}}} \right\rangle  = {\cal N}\int_{{t_N} >  \ldots  > {t_1}} {\left[ {g\left( {{t_1},{t_2}, \ldots ,{t_N}} \right)\prod\limits_{i = 1}^N {a_{in}^\dag \left( {{t_i}} \right)d{t_i}} } \right]} \left| 0 \right\rangle
    \end{eqnarray}
    
In the above expression, $g$ is the $N$-photon amplitude function and  is the normalization factor arising from the indistinguishability and sorting of the photons. Additionally, the $N$-fold integral used is defined as follows:
\begin{eqnarray}
	\int_{{t_N} >  \ldots  > {t_1}} {d{t_N}d{t_{N - 1}} \ldots d{t_1}} : = \int_{ - \infty }^\infty  {d{t_N}} \int_{ - \infty }^{{t_N}} {d{t_{N - 1}}}  \ldots \int_{ - \infty }^{{t_2}} {d{t_1}}
    \end{eqnarray}
    
In our analysis, we consider various input types based on time correlations and the number of wavepackets. We begin with a single-wavepacket Gaussian-shaped pulsed input, for which the amplitude function is expressed as:
\begin{eqnarray}
	g\left( {{t_1},{t_2}, \ldots ,{t_N}} \right) = \prod\limits_{i = 1}^N {h\left( {{t_i}} \right)}
\end{eqnarray}
with $h\left( t \right) = \sqrt {{2 \mathord{\left/
 {\vphantom {2 {(\delta \sqrt \pi  )}}} \right.
 \kern-\nulldelimiterspace} {(\delta \sqrt \pi  )}}} \exp \left( { - i{\omega _0}t - {{2{t^2}} \mathord{\left/
 {\vphantom {{2{t^2}} {{\delta ^2}}}} \right.
 \kern-\nulldelimiterspace} {{\delta ^2}}}} \right)$ being the single-photon amplitude function of the Gaussian wavepacket with width $\delta $ and carrier frequency ${\omega _0}$. The normalization factor in this case is given by $\mathcal{N}=\sqrt{N!}$  \cite{RN106}.

 \begin{figure}
\includegraphics{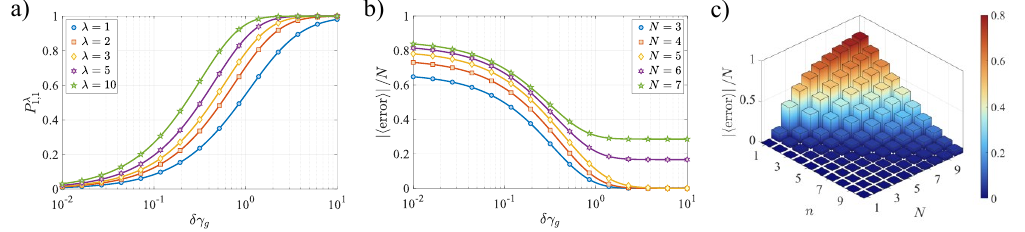}
\caption{\label{fig:epsart} a) Probability of successfully subtracting a single photon by a cascade of $\lambda $ emitters, based on equation (16), b) Relative error of a PNR detector with a cascade of $n = 4$ emitters, for input pulses with different photon numbers $N$, and c) Relative error of a PNR detector as a function of the input photon number $N$ and the number of emitters $n$ for $\delta {\gamma _g} = 0.5$.
}
\end{figure}

\subsection{Linear analysis}

In the linear regime, the system's behavior can be fully predicted through a single-photon transport analysis. In this case, the transfer function of each photon subtractor between its input and each of its two outputs can be determined by taking the Fourier transform of the scattering matrix elements $\Sigma _1^1\left( {t' - t} \right)$ and $\Sigma _0^1\left( {t' - t} \right)$ given in equation (8), yielding the well-known Lorentzian-shaped frequency responses $\Sigma _1^1\left( \omega  \right) = {{ - {\gamma _g}} \mathord{\left/
 {\vphantom {{ - {\gamma _g}} {\left( {{\gamma _g} + i\omega } \right)}}} \right.
 \kern-\nulldelimiterspace} {\left( {{\gamma _g} + i\omega } \right)}}$ and $\Sigma _0^1\left( \omega  \right) = {{i\omega } \mathord{\left/
 {\vphantom {{i\omega } {\left( {{\gamma _g} + i\omega } \right)}}} \right.
 \kern-\nulldelimiterspace} {\left( {{\gamma _g} + i\omega } \right)}}$\cite{RN173}, where $\omega$ is the detuning from the atomic resonance frequency ${\omega _{13}}$. Using these transfer functions, we can calculate the probability that a single photon present at the input is subtracted by a cascade of $\lambda $ emitters (each in their $\left| 1 \right\rangle $-state), as:
 \begin{eqnarray}
	P_{1,1}^\lambda  = 1 - \int {{{\left| {{{\left( {\Sigma _0^1\left( \omega  \right)} \right)}^\lambda }h\left( {\omega  + {\omega _{13}}} \right)} \right|}^2}d\omega }
    \end{eqnarray}
where $h\left( {\omega '} \right) = \sqrt {{\delta  \mathord{\left/
 {\vphantom {\delta  {(2\sqrt \pi  )}}} \right.
 \kern-\nulldelimiterspace} {(2\sqrt \pi  )}}} \exp \left( { - {{{{\left( {\omega ' - {\omega _0}} \right)}^2}{\delta ^2}} \mathord{\left/
 {\vphantom {{{{\left( {\omega ' - {\omega _0}} \right)}^2}{\delta ^2}} 8}} \right.
 \kern-\nulldelimiterspace} 8}} \right)$ is the Fourier transform of $h\left( t \right)$, representing the frequency distribution of each single photon. An important point regarding the above expression is that when the single-photon pulse passes through each emitter without being subtracted, it is reshaped according to $\Sigma _0^1\left( \omega  \right)$, which reduces its likelihood of being absorbed by subsequent emitters. If we have a PNR detector with $n$ emitters, as shown in Fig. 1.b, we can use equation (16) to calculate the probability of successfully subtracting $k$ photons out of $N$ photons at the input, denoted by $P_{N,k}^n$. This is determined by summing the probabilities of successfully subtracting all possible combinations of $k$ photons out of $N$ photons, while accounting for the fact that after each successful subtraction, the number of active emitters decreases by one, resulting in:
 \begin{eqnarray}
	P_{N,k}^n = \left\{ \begin{array}{l}
\left( {\prod\limits_{m = 0}^{k - 1} {P_{1,1}^{n - m}} } \right)\sum\limits_{B_{N - k}^k} {\left( {\prod\limits_{i = 1}^{N - k} {\left( {1 - P_{1,1}^{n - B_{N - k}^k\left( i \right)}} \right)} } \right)} \,\,\,\,\,\,\,\,\,\,\,\,\,\,\,\,\,k > 0\\
{\left( {1 - P_{1,1}^n} \right)^N}\,\,\,\,\,\,\,\,\,\,\,\,\,\,\,\,\,\,\,\,\,\,\,\,\,\,\,\,\,\,\,\,\,\,\,\,\,\,\,\,\,\,\,\,\,\,\,\,\,\,\,\,\,\,\,\,\,\,\,\,\,\,\,\,\,\,\,\,k = 0
\end{array} \right.
\end{eqnarray}
where $B_{N - k}^k$ is an unordered set with $N - k$ elements chosen from $\left\{ {0,1, \ldots ,k} \right\}$. If $k = N$ or $k = N - 1$, the number of detectors that register a click equals $N$, resulting in accurate photon counting. Otherwise, there is an error of $k + 1 - N$ in counting the photons. Therefore, the average error of the detector in counting the input photons can be written as:
\begin{eqnarray}
	\left\langle {{\rm{error}}} \right\rangle  = \sum\limits_{k = 0}^{\min \left( {n,N - 2} \right)} {\left( {k + 1 - N} \right)P_{N,k}^n}
    \end{eqnarray}

If in an $n$-emitter PNR detector subjected to an $n + 1$ photon pulse any emitter fails to subtract a photon, the resulting photon count will be inaccurate. This occurs because multiple photons may arrive at the last detector, which lacks photon-number-resolving capability and, regardless of the number of arriving photons, registers only a single click. One approach to increase the precision of the detector is to add additional emitters. This way, if one emitter fails to subtract a photon, other emitters downstream can compensate by subtracting it, ensuring the final count remains accurate. However, as mentioned above, the pulse reshaping occurred during its interaction with upstream emitters makes it harder for downstream emitters to subtract a photon. Fig. 2 shows the results for the linear regime based on equations (16) to (18) when the carrier frequency of the input pulses is in resonance with the atomic transition, i.e., ${\omega _0} = {\omega _{13}}$. Parts (b) and (c) are particularly relevant from a design perspective, as they provide insight into the number of emitters required to count photons in the input pulse within a specified error margin. The results are plotted as functions of $\delta {\gamma _g}$ representing the ratio of the atomic linewidth to the spectral bandwidth of the input pulse.

\subsection{Nonlinear analysis}

In the nonlinear regime, in addition to the pulse shaping discussed above, upstream emitters can also influence the probability of photon subtraction by downstream counterparts through a second mechanism: generating correlations between their scattered photons by forming multi-photon bound states \cite{RN175, RN176}. To investigate this correlation, we first define the zero-delay second-order correlation function ${g^{\left( 2 \right)}}\left( 0 \right)$ for photons exiting the photon subtractor through the ${a_{out,1}}$ output mode (see Fig. 1.a). By averaging and normalizing the two-time correlator ${G^{\left( 2 \right)}}\left( {{t_1},{t_2}} \right)$, the following expression for ${g^{\left( 2 \right)}}\left( 0 \right)$ is derived for a pulse containing $N$ photons:
\begin{eqnarray}
	{g^{\left( 2 \right)}}\left( 0 \right) = {\left( {N - 1} \right)^2}\frac{{\int {{G^{\left( 2 \right)}}\left( {t,t} \right)dt} }}{{\int {{{\left| {\int {{G^{\left( 2 \right)}}\left( {t,t'} \right)dt'} } \right|}^2}dt} }}
    \end{eqnarray}

In the derivation of the above expression, the following relation between the first-order and second-order correlation functions for an $N$-photon pulse is used:
\begin{eqnarray}
	{{G}^{\left( 1 \right)}}\left( t,t \right)=\frac{1}{N-1}\int{{{G}^{\left( 2 \right)}}\left( t,{t}' \right)d{t}'}
    \end{eqnarray}
 
Generation of photon-photon correlations at the ${a_{out,1}}$ output requires that at least two photons in this output have interacted with the emitter. Since the emitter's interaction with the input photons ceases as soon as a photon is subtracted, such correlations can occur only if no photon is subtracted from an input pulse with at least two photons, or if the third photon or any subsequent photon is subtracted from input pulses with at least three photons. Consequently, here, we study three specific cases: first, when two photons are present at the input and no photon is subtracted, second, when three photons are present at the input and no photon is subtracted, and third, when three photons are present at the input and the third photon is successfully subtracted. The derived scattering matrix elements for two- and three-photon inputs are provided in the Appendix. As we will show, this analysis provides valuable insights into the behavior observed during nonlinear $N$-photon transport through a cascaded system.

\begin{figure}
\includegraphics{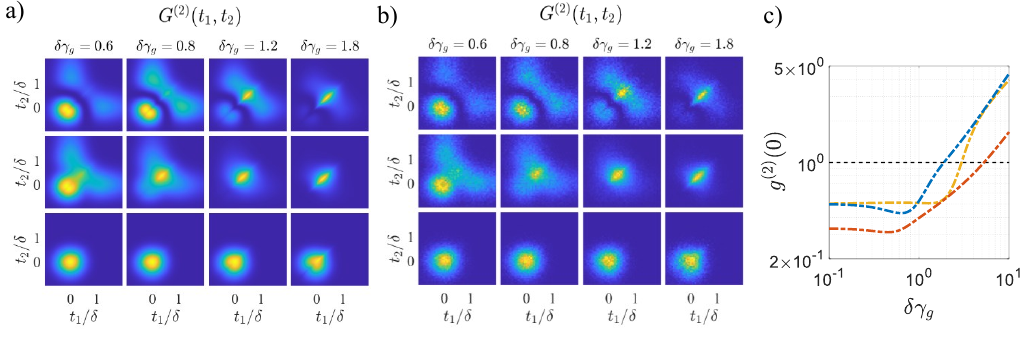}
\caption{\label{fig:epsart} a) Second-order correlation of the output photons at the ${a_{out,1}}$ mode of a single-emitter photon subtractor, shown for various values of $\delta {\gamma _g}$. The first row corresponds to a two-photon input with no photon subtracted, the second row to a three-photon input with no photon subtracted, and the third row to a three-photon input where the third photon is subtracted. b) Same results as in (a), obtained using the Monte-Carlo method by averaging across 50,000 trajectories. c) Zero-delay single-time correlation function ${g^{\left( 2 \right)}}\left( 0 \right)$ as a function of $\delta {\gamma _g}$ for the three cases mentioned above, represented by the blue, red, and yellow curves, respectively.
}
\end{figure}

Using the derived scattering matrix elements given in equation (9), the amplitude function of photons after interacting with a single emitter, for the case where the $j$-th photon is subtracted (denoted as ${g'_{N,j}}\left( {{{t'}_1}, \ldots ,{{t'}_N}} \right)$), can be expressed as:
\begin{eqnarray}
{{{g}'}_{N,j}}\left( {{{{t}'}}_{1}},\ldots ,{{{{t}'}}_{N}} \right)=\int_{{{{{t}'}}_{N-1}}}^{{{{{t}'}}_{N}}}{d{{t}_{N}}\ldots \int_{{{{{t}'}}_{1}}}^{{{{{t}'}}_{2}}}{d{{t}_{2}}\int_{-\infty }^{{{{{t}'}}_{1}}}{d{{t}_{1}}\left[ \sum\limits_{per{{m}_{N}}}{\Sigma _{j}^{N}\left( {{t}_{1}},\ldots ,{{t}_{N}};{{{{t}'}}_{per{{m}_{N}}\left( 1 \right)}},\ldots ,{{{{t}'}}_{per{{m}_{N}}\left( N \right)}} \right)} \right]g\left( {{t}_{1}},\ldots ,{{t}_{N}} \right)}}}\nonumber
\end{eqnarray}\begin{eqnarray}\end{eqnarray}
where $per{m_N}$ denotes a permutation of numbers $1$ to $N$. This amplitude function can be used to derive the second-order correlation of the output photons in the form:
\begin{eqnarray}
	{G^{\left( 2 \right)}}\left( {{{t'}_1},{{t'}_2}} \right) = N\left( {N - 1} \right)\int {{{\left| {{{g'}_{N,j}}\left( {{{t'}_1}, \ldots ,{{t'}_N}} \right)} \right|}^2}d{{t'}_3} \ldots d{{t'}_N}}
\end{eqnarray}

In addition, the probability that the $j$-th photon is subtracted from an $N$-photon input pulse is given by:
\begin{eqnarray}
	P_j^N = \int {{{\left| {{{g'}_{N,j}}\left( {{{t'}_1}, \ldots ,{{t'}_N}} \right)} \right|}^2}d{{t'}_1} \ldots d{{t'}_N}}
\end{eqnarray}

\begin{figure}
\includegraphics{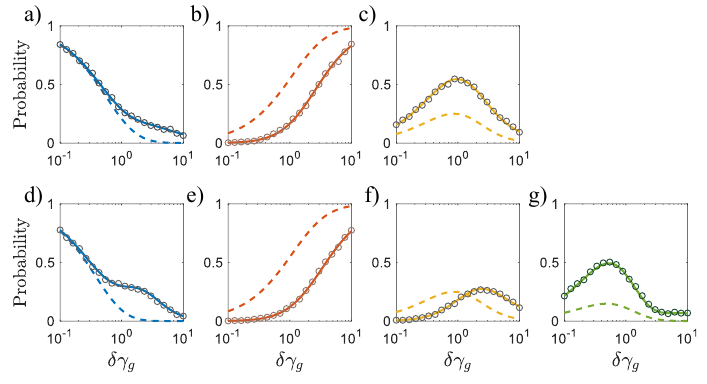}
\caption{\label{fig:epsart} (a, b, c) Probabilities of subtracting no photon, the first photon, and the second photon from a two-photon input, respectively. (d, e, f, g) Probabilities of subtracting no photon, the first photon, the second photon, and the third photon from a three-photon input, respectively. The solid and dashed lines represent the nonlinear and linear analytical solutions, respectively, while the circle markers denote the numerical solutions obtained using the quantum trajectory method, averaged across 2000 trajectories.
}
\end{figure}

Fig. 3 shows the second-order autocorrelation function of the output photons for the cases described below equation (20). Each plot in parts (a) and (b) of the figure is normalized to its maximum value for improved visual clarity. As can be seen, in all parts of the figure, the output light in all three cases exhibits some degree of bunching in the strong interaction regime, where $\delta {\gamma _g} > 1$. This phenomenon arises from the enhanced two-photon transition due to stimulated emission \cite{RN177}. The generated anti-bunching observed in the weak interaction regime with $\delta {\gamma _g} < 1$ arises when one of the two incident photons is absorbed by the emitter, delaying its re-emission, while the second photon simply passes through without interacting. The numerical results are obtained through simulations of the quantum network described in section 2.3 using the quantum trajectory method with the QuTiP toolbox in Python, showing good agreement with their analytical counterparts.

In Fig. 4, for the two cases of two and three input photons, the probabilities of different outcomes are presented, including the possibilities of subtracting no photon or subtracting each of the contained photons in the input pulse based on their arrival order. The nonlinear solutions are calculated from equation (23). For the linear solution, the probability of subtracting no photon out of $N$ photons is given by ${\left( {1 - P_{1,1}^1} \right)^N}$, and the probability of subtracting the $j$-th photon is given by ${\left( {1 - P_{1,1}^1} \right)^{j - 1}}P_{1,1}^1$, where $P_{1,1}^1$ is determined using Equation (16). As seen in parts (a) and (d) of the figure, in the weak interaction regime where $\delta {\gamma _g} < 1$, the linear model can predict the failure probability of the process—resulting in no photon being subtracted—with reasonable precision. However, as shown in the other parts of the figure, the linear solution cannot reliably determine which photon will be subtracted. This occurs due to nonlinear scattering induced by atomic saturation, leading to the reordering of the scattered photons. The first arriving photon excites and saturates the atom but is routed to the other waveguide mode only after the second photon has passed. Another essential observation from this figure is that in the strong coupling regime where the probability of subtracting no photon approaches zero, the first photon is almost deterministically subtracted, as predicted by both linear and nonlinear models. This process temporally entangles the subtracted photon with the remaining photons, ensuring that the subtracted photon is always detected before the remaining photons, a phenomenon that has been experimentally confirmed by Rosenblum et al. \cite{RN36}. The higher failure rate of photon subtraction predicted by the nonlinear model arises from the stimulated emission effect when multiple photons interact with the atom within its response time. This interaction leads to the formation of bound states at the ${a_{out,1}}$ output, which is not accounted for in the linear model.

\begin{figure}
\includegraphics{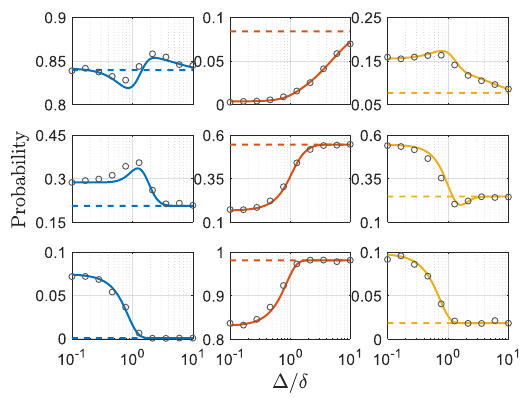}
\caption{\label{fig:epsart} (a, b, c) Probabilities of subtracting no photon, the first photon, and the second photon from a two-photon input, respectively. (d, e, f, g) Probabilities of subtracting no photon, the first photon, the second photon, and the third photon from a three-photon input, respectively. The solid and dashed lines represent the nonlinear and linear analytical solutions, respectively, while the circle markers denote the numerical solutions obtained using the quantum trajectory method, averaged across 2000 trajectories.
}
\end{figure}

So far, we have investigated the performance of SPRINT-based photon subtractors for inputs with different photon numbers, all of which were in a single temporal mode. We observed that the width of the input pulses, relative to the atomic response time, plays a key role in the successful subtraction of photons. Now, we will explore the system's response to inputs with multiple modes. First, to examine how the photons’ overlap affects the validity of the linear solution, let us consider a two-photon state consisting of two single-photon Gaussian wavepackets separated by a time interval $\Delta $. The symmetrized amplitude function for this state will be:
\begin{eqnarray}
	g\left( {{t_1},{t_2}, \ldots ,{t_N}} \right) = \sum\limits_{per{m_N}} {\left[ {\prod\limits_{i = 1}^N {h\left( {{t_{per{m_N}\left( i \right)}} - i\Delta } \right)} } \right]}
    \end{eqnarray}

In this case, the photons have partial indistinguishability, and the normalization factor is , where ${\cal S}$ is the indistinguishability matrix, with its elements given by the inner product of different wavepackets \cite{RN160}, i.e., ${{\cal S}_{ij}} = \left\langle {{h(t - i\Delta )}}
 \mathrel{\left | {\vphantom {{h(t - i\Delta )} {h(t - j\Delta )}}}
 \right. \kern-\nulldelimiterspace}
 {{h(t - j\Delta )}} \right\rangle $. For the two-photon input with Gaussian wavepackets introduced earlier, this results in . By structuring the input in this way, it becomes possible to investigate the system’s response in both linear and nonlinear regimes simply by adjusting the parameter $\Delta $. When $\Delta  \gg \delta $ and $\Delta  \gg {{1 \mathord{\left/
 {\vphantom {1 \gamma }} \right.
 \kern-\nulldelimiterspace} \gamma }_g}$, only one photon interacts with the system at a time, resulting in a linear response. For smaller values of $\Delta $, two-photon effects may arise, leading to a nonlinear response. In the limit of $\Delta  = 0$, the input resembles an $2$-photon pulse within a single wavepacket, as considered before. The results shown in Fig. 5 correspond to the case of two-photon input. As $\Delta $ increases, leading to a lower overlap between the photons, the results converge to the linear solution. Additionally, as seen in part (a) of the figure, in the weak interaction regime ($\delta {\gamma _g} \ll 1$), the linear solution for the success probability of the photon subtractor remains valid with high precision, regardless of the amount of overlap. This is because, in this regime, the probability of two-photon interactions remains low, even when photons arrive at the atom within a short time interval.
 
To further verify the accuracy of our analytical and numerical models for inputs with more than one mode, particularly those that exhibit correlations or further anti-correlations compared to Fock-state inputs, we consider the following two-mode inputs: 
	\begin{eqnarray}
    \begin{array}{*{20}{c}}
{{g_1}\left( {{t_1},{t_2}} \right) = {h_0}\left( {{t_1}} \right){h_0}\left( {{t_2}} \right) + {h_1}\left( {{t_1}} \right){h_1}\left( {{t_2}} \right)}&{\,\,\,\,\,\,\,\,\,\,(a}\\
{{g_2}\left( {{t_1},{t_2}} \right) = {h_0}\left( {{t_1}} \right){h_0}\left( {{t_2}} \right) - {h_1}\left( {{t_1}} \right){h_1}\left( {{t_2}} \right)}&{\,\,\,\,\,\,\,\,\,\,(b}
\end{array}
\end{eqnarray}
where ${h_0}\left( t \right)$ and ${h_1}\left( t \right)$ are the zero-th and first-order 1D Hermite-Gauss functions of width $\delta $, respectively, which are defined generally for the $n$-th order as ${h_n}\left( t \right) = \sqrt {{1 \mathord{\left/
 {\vphantom {1 {{2^{n - 1}}n!\sqrt \pi  \delta }}} \right.
 \kern-\nulldelimiterspace} {{2^{n - 1}}n!\sqrt \pi  \delta }}} \exp \left( { - 2{{{t^2}} \mathord{\left/
 {\vphantom {{{t^2}} {{\delta ^2}}}} \right.
 \kern-\nulldelimiterspace} {{\delta ^2}}}} \right){{\rm{H}}_n}\left( {{{2t} \mathord{\left/
 {\vphantom {{2t} \delta }} \right.
 \kern-\nulldelimiterspace} \delta }} \right)$, with ${{\rm{H}}_n}\left( t \right)$ being the $n$-th order Hermite polynomial. As seen in the first column of Figs. 6.a and 6.b, the amplitude functions given in Eq. (25) generate bunched and anti-bunched light, respectively, compared to the single-mode two-photon Fock state. The correlation between the output photons, in cases where no photon is subtracted by the photon subtractor, is shown in the subsequent columns of Figs. 6.a and 6.b, as well as in Fig. 6.c. As shown in this figure, bound states dominate the output in the strong interaction regime for both input states. Additionally, in the intermediate interaction regime ($\delta {\gamma _g} \sim 1$), the bunched input exhibits stronger anti-bunching. This occurs because, when photons arrive close together, the probability of only one photon interacting with the atom increases, while the other photon is directly transmitted due to atomic saturation. In Fig. 6(d), the probabilities of subtracting no photon, the first photon, and the second photon are plotted for these input pulses. For the bunched input, there is a higher probability that the subtracted photon is detected after the transmitted photon due to the small arrival-time difference between the photons (see the rightmost plot). Additionally, the total success rate of subtraction is lower for both bunched and anti-bunched inputs compared to the single-mode Fock state input (see the leftmost plot). This occurs because, in the anti-bunched case, the effective interaction time of each photon with the emitter is shorter. In the bunched case, the reduction arises because the second photon arrives shortly after the first, decreasing its likelihood of interacting with the emitter due to saturation.

\begin{figure}
\includegraphics{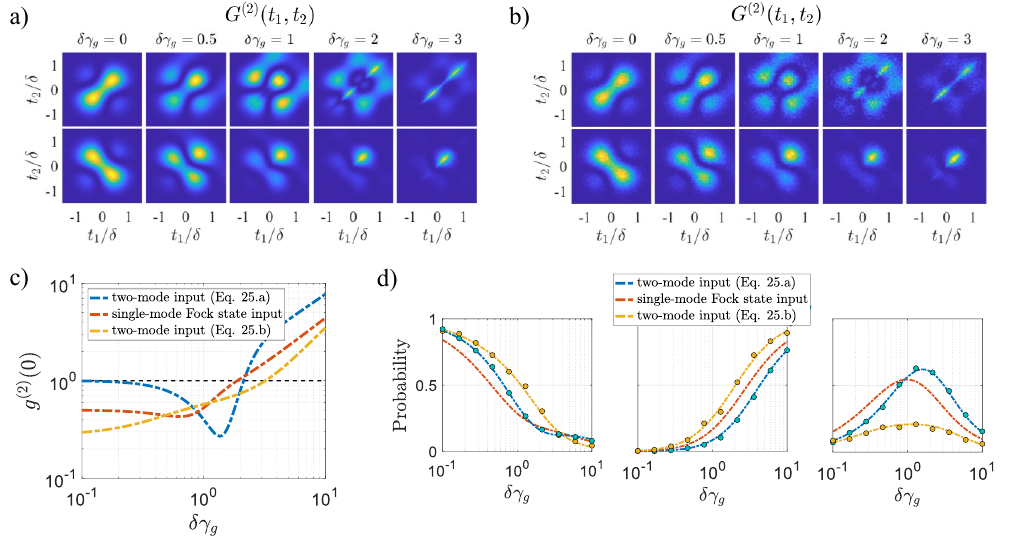}
\caption{\label{fig:epsart} a) Second-order correlation of the output photons at the ${a_{out,1}}$ mode of a single-emitter photon subtractor, shown for various values of $\delta {\gamma _g}$. The first and second rows corresponds to a bunched and anti-bunched two-photon input with amplitudes given in equations (25a) and (25b), respectively. b) Same results as in (a), obtained using the Monte-Carlo method by averaging across 50,000 trajectories. c) Zero-delay single-time correlation function ${g^{\left( 2 \right)}}\left( 0 \right)$ of the output light as a function of $\delta {\gamma _g}$ for these two correlated input states, along with the uncorrelated input, as described in equation (15). d, e, f) Probability of successfully subtracting a single photon, the first arriving photon, and the second arriving photon, respectively. The circle markers denote the numerical solutions obtained using the quantum trajectory method, averaged across 1,000 trajectories.
}
\end{figure}

\begin{figure}
\includegraphics{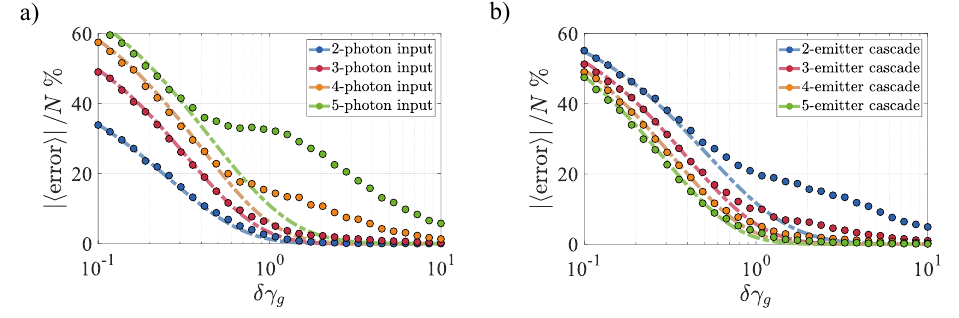}
\caption{\label{fig:epsart} a) Relative error of a PNR detector consisting of a cascade of $n = 4$ quantum emitters when subjected to Gaussian input pulses with different photon numbers. b) Relative error of PNR detectors with cascades of $n = 2$ to $n = 5$ quantum emitters when subjected to Gaussian input pulses with $N = 3$ photons. The dash-dotted lines indicate the results from the linear analysis, while the dots represent nonlinear results calculated using the quantum trajectory method, averaged across 5,000 trajectories.
}
\end{figure}

As we observed in the previous results, the stimulated emission effect in the nonlinear regime prevents single emitters from successfully subtracting photons by inducing the re-emission of the absorbed photon into the same waveguide mode. Additionally, it generates bound states that due to their bunching are more difficult to be re-routed by the next emitter in the cascade. Therefore, we expect the precision of a PNR detector to be lower than that predicted by the linear model (see Fig. 2) in the region where these bound states are generated. However, increasing the number of photon subtractors in the cascade relative to the number of input photons can enhance this precision, leading to highly accurate detectors, particularly in the strong interaction regime. This is confirmed by Fig. 7, which shows the relative error of a detector with different number of emitters and input photons as predicted by both linear and nonlinear models.

To evaluate the photon-number-resolving performance of the proposed detector, we compare it against conventional PNR techniques that rely on linear optical splitting of the input pulse across multiple single-photon detectors. In these approaches, the photons are distributed across several detector pixels, with the ideal case being a uniform distribution. The total photon number is then estimated by counting the number of detectors that register a click. When the number of detector pixels significantly exceeds the number of incident photons, the probability of two photons arriving at the same detector becomes negligible, enabling accurate photon counting. Spatial demultiplexing of photons in these methods is typically achieved using balanced beamsplitter trees \cite{RN233}, cascaded polarizing beamsplitters and half-wave plates \cite{RN196}, multipixel sensor grids \cite{RN232}, or cascades of waveguide-coupled superconducting nanowire single-photon detectors (SNSPDs) \cite{RN231}. Fig. 8a illustrates a typical setup consisting of a cascade of beamsplitters with reflectivities ${R_1}$ to ${R_n}$, where each beamsplitter reflects a portion of the incident light to a single-photon detector, and the final detector collects the fully transmitted light. The probability that an incoming photon is routed to the $i$ th detector in this setup is given by:
\begin{eqnarray}
	{P_i} = \left\{ \begin{array}{l}
{R_i}\,\,\,\,\,\,\,\,\,\,\,\,\,\,\,\,\,\,\,\,\,\,\,\,\,\,\,\,\,\,\,\,\,\,\,\,\,\,\,\,\,\,\,\,\,i = 1\\
{R_i}\prod\nolimits_{j = 1}^{i - 1} {\left( {1 - {R_j}} \right)} \,\,\,\,\,\,\,\,\,\,\,\,\,\,\,\,1 < i < n + 1\\
\prod\nolimits_{j = 1}^{i - 1} {\left( {1 - {R_j}} \right)} \,\,\,\,\,\,\,\,\,\,\,\,\,\,\,\,\,\,\,\,\,i = n + 1
\end{array} \right.
\end{eqnarray}

To describe the probability that $k$ detectors out of $n + 1$ register a click when $N$ photons are sent into the input of the beamsplitter cascade, we model the system as a multinomial distribution over the $n + 1$ detectors, where each detector $i$ is associated with a photon-routing probability ${P_i}$. This leads to the following expression for the average number of clicks registered by the detectors:
\begin{eqnarray}
	\left\langle {{\rm{clicks}}} \right\rangle  = N!\sum\limits_{k = 0}^{\min \left( {n + 1,N} \right)} {\sum\limits_{A_{n + 1}^{N,k}} {\left[ {{{\left( {\prod\limits_{i = 1}^{n + 1} {P_i^{A_{n + 1}^{N,k}\left( i \right)}} } \right)} \mathord{\left/
 {\vphantom {{\left( {\prod\limits_{i = 1}^{n + 1} {P_i^{A_{n + 1}^{N,k}\left( i \right)}} } \right)} {\left( {\prod\limits_{i = 1}^{n + 1} {A_{n + 1}^{N,k}\left( i \right)!} } \right)}}} \right.
 \kern-\nulldelimiterspace} {\left( {\prod\limits_{i = 1}^{n + 1} {A_{n + 1}^{N,k}\left( i \right)!} } \right)}}} \right]} }
 \end{eqnarray}
where $A_{n + 1}^{N,k}$ is an ordered set of $n + 1$ non-negative integers whose sum is $N$, and exactly $k$ of the elements are nonzero. 
Fig. 8b shows the detector response curve, which plots the average number of detector clicks as a function of the number of incident photons for an input pulse prepared in a Fock state with photon number $N$. The blue curve represents the most accurate response achievable with the conventional scheme shown in Fig. 8a. This optimal response is obtained when the beamsplitter reflectivities are configured such that photons are uniformly distributed among the detectors—that is, when ${P_1} = {P_2} =  \ldots  = {P_{n + 1}}$, as in the case of a balanced beamsplitter tree. The orange curve corresponds to the response of the same conventional scheme, but with beamsplitters of equal reflectivity—that is, when, ${R_1} = {R_2} =  \ldots  = {R_n}$, which is equivalent to the case of a PNR detector using a series of SNSPDs placed along a waveguide \cite{RN231}. In this case, the reflectivity value is chosen to maximize the precision of the detector. The other curves correspond to the response of the proposed detector for various values of the parameter $\delta {\gamma _g}$. Both approaches are assumed to use the same number of single-photon detectors, i.e., five. As shown, the detector proposed in this paper outperforms the conventional scheme when $\delta {\gamma _g}$ reaches values on the order of approximately 10. Assuming an input pulse width of 10 ns - which corresponds to the typical dead time of currently available single-photon detectors based on avalanche photodiodes or SNSPDs - this condition is satisfied when the coupling rate ${\gamma _g}$ reaches $1\,{\rm{GHz}}$. At the same time, coupling rates of several gigahertz have been reported for solid-state emitters coupled to chiral photonic crystal waveguides \cite{RN139, RN234}, which highlight the feasibility of implementing the proposed approach for high-precision photon-number-resolving detection with currently available technologies.

\begin{figure}
\includegraphics{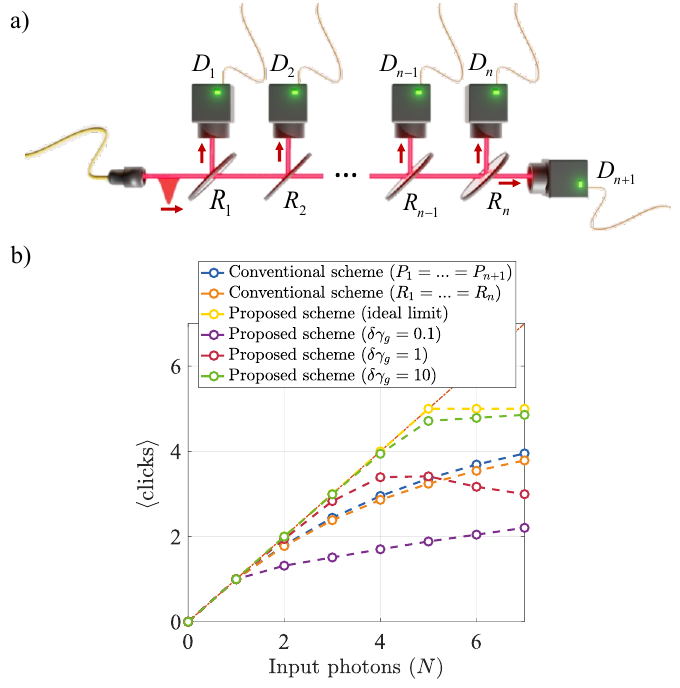}
\caption{\label{fig:epsart} a) Schematic of a conventional PNR detector, which spatially distributes input photons among multiple single-photon detectors using a cascade of beamsplitters. b) Performance comparison between the most accurate conventional scheme achievable with $n = 4$ beamsplitters and the photon-subtraction-based scheme with a cascade of $n = 4$ quantum emitters, shown for different values of the parameter $\delta {\gamma _g}$. The results for the proposed scheme are calculated using the quantum trajectory method, averaged across 5,000 trajectories.
}
\end{figure}

\section{Conclusion and outlook}
In summary, we proposed and investigated in detail a PNR detector based on a cascade of $\Lambda$-type emitters coupled to a chiral waveguide. In this potential application of SPRINT-based deterministic photon subtraction, the cascade of waveguide-coupled $\Lambda$-type atoms enables the deterministic demultiplexing of input photons among multiple single-photon detectors. Based on a photon transport analysis in the time domain, we calculated the success rate of photon subtraction for input pulses with varying photon numbers, considering single-mode pulses, as well as both fully or partially overlapping  multi-mode pulses. We further evaluated the total number of photons subtracted by the cascade and determined the achievable resolution precision as a function of the number of emitters and their coupling strength to the waveguide. Our results show that the proposed detector, with technologically feasible coupling rates, can outperform conventional spatial demultiplexing PNR schemes using the same number of single-photon detectors. In our future work, we plan to further explore how this PNR approach compares to the conventional demultiplexing PNR as the number of detectors scales beyond the few-photon case-studies presented here and for detection of more complex non-classical states. 
At the same time, our PNR detector offers several unusual features, such as the generation of correlations in the nonlinear regime and intermittent entanglement between photons during the detection process. These are not available in the conventional demultiplexing PNR schemes and open new possibilities for use in fundamental studies and practical applications. In addition to detecting multi-photon states with high precision, one can envision this PNR to be used for novel forms of quantum tomography and for generation of exotic photon states by combining the PNR and entanglement generation capabilities of the cascaded photon subtraction. To look for inspiration, one can consider past studies in which primitives of photon addition and subtraction were employed for detection and generation of quantum states of light  \cite{RN191, RN193, RN198, RN192, RN200, RN41}. For example, multi-photon subtraction can extend our range of non-classical states \cite{RN190, RN41} that can be generated from nowadays standard sources of light such as coherent or squeezed light sources. Combining both these capabilities of the system would then allow for ‘number-resolved’ subtraction \cite{RN199, RN200, RN201} which can potentially increase the generation rate of useful nonclassical states.

\begin{acknowledgments}
This research was undertaken thanks in part to funding from the Canada First Research Excellence Fund Transformative Quantum Technologies program (CFREF-TQT).
\end{acknowledgments}

\section*{appendix}
    Using equations (9) and (10), the scattering matrix elements for the two cases of two- and three-photon inputs can be written as follows:
\begin{eqnarray}
\begin{array}{l}
\Sigma _1^2\left( {{t_1},{t_2};{{t'}_1},{{t'}_2}} \right) = \Sigma _1^1\left( {{{t'}_1} - {t_1}} \right)\delta \left( {{t_2} - {{t'}_2}} \right)\\
\Sigma _2^2\left( {{t_1},{t_2};{{t'}_1},{{t'}_2}} \right) = \Sigma _0^1\left( {{{t'}_1} - {t_1}} \right)\Sigma _1^1\left( {{{t'}_2} - {t_2}} \right) - {\gamma _g}\delta \left( {{t_2} - {{t'}_1}} \right)\exp \left( {{\gamma _g}\left( {{t_1} - {{t'}_2}} \right)} \right)\\
\Sigma _0^2\left( {{t_1},{t_2};{{t'}_1},{{t'}_2}} \right) = \Sigma _0^1\left( {{{t'}_1} - {t_1}} \right)\Sigma _0^1\left( {{{t'}_2} - {t_2}} \right) - {\gamma _g}\delta \left( {{t_2} - {{t'}_1}} \right)\exp \left( {{\gamma _g}\left( {{t_1} - {{t'}_2}} \right)} \right)\\
\Sigma _1^3\left( {{t_1},{t_2},{t_3};{{t'}_1},{{t'}_2},{{t'}_3}} \right) = \Sigma _1^1\left( {{{t'}_1} - {t_1}} \right)\delta \left( {{t_2} - {{t'}_2}} \right)\delta \left( {{t_3} - {{t'}_3}} \right)\\
\Sigma _2^3\left( {{t_1},{t_2},{t_3};{{t'}_1},{{t'}_2},{{t'}_3}} \right) = \Sigma _0^1\left( {{{t'}_1} - {t_1}} \right)\Sigma _1^1\left( {{{t'}_2} - {t_2}} \right)\delta \left( {{t_3} - {{t'}_3}} \right) - {\gamma _g}\delta \left( {{t_2} - {{t'}_1}} \right)\delta \left( {{t_3} - {{t'}_3}} \right)\exp \left( {{\gamma _g}\left( {{t_1} - {{t'}_2}} \right)} \right)\\
\Sigma _3^3\left( {{t_1},{t_2},{t_3};{{t'}_1},{{t'}_2},{{t'}_3}} \right) = \Sigma _0^1\left( {{{t'}_1} - {t_1}} \right)\Sigma _0^1\left( {{{t'}_2} - {t_2}} \right)\Sigma _1^1\left( {{{t'}_3} - {t_3}} \right) - {\gamma _g}\delta \left( {{t_2} - {{t'}_1}} \right)\Sigma _1^1\left( {{{t'}_3} - {t_3}} \right)\exp \left( {{\gamma _g}\left( {{t_1} - {{t'}_2}} \right)} \right)\\
\,\,\,\,\,\,\,\,\,\,\,\,\,\,\,\,\,\,\,\,\,\,\,\,\,\,\,\,\,\,\,\,\,\,\,\,\,\,\,\,\,\,\,\,\,\, - {\gamma _g}\Sigma _0^1\left( {{{t'}_1} - {t_1}} \right)\delta \left( {{t_3} - {{t'}_2}} \right)\exp \left( {{\gamma _g}\left( {{t_2} - {{t'}_3}} \right)} \right) - {\gamma _g}\delta \left( {{t_2} - {{t'}_1}} \right)\delta \left( {{t_3} - {{t'}_2}} \right)\exp \left( {{\gamma _g}\left( {{t_1} - {{t'}_3}} \right)} \right)\\
\Sigma _0^3\left( {{t_1},{t_2},{t_3};{{t'}_1},{{t'}_2},{{t'}_3}} \right) = \Sigma _0^1\left( {{{t'}_1} - {t_1}} \right)\Sigma _0^1\left( {{{t'}_2} - {t_2}} \right)\Sigma _0^1\left( {{{t'}_3} - {t_3}} \right) - {\gamma _g}\delta \left( {{t_2} - {{t'}_1}} \right)\Sigma _0^1\left( {{{t'}_3} - {t_3}} \right)\exp \left( {{\gamma _g}\left( {{t_1} - {{t'}_2}} \right)} \right)\\
\,\,\,\,\,\,\,\,\,\,\,\,\,\,\,\,\,\,\,\,\,\,\,\,\,\,\,\,\,\,\,\,\,\,\,\,\,\,\,\,\,\,\,\,\,\, - {\gamma _g}\Sigma _0^1\left( {{{t'}_1} - {t_1}} \right)\delta \left( {{t_3} - {{t'}_2}} \right)\exp \left( {{\gamma _g}\left( {{t_2} - {{t'}_3}} \right)} \right) - {\gamma _g}\delta \left( {{t_2} - {{t'}_1}} \right)\delta \left( {{t_3} - {{t'}_2}} \right)\exp \left( {{\gamma _g}\left( {{t_1} - {{t'}_3}} \right)} \right)
\\
\\
\end{array}\nonumber
\end{eqnarray}

\bibliography{apssamp}

\end{document}